\documentclass[12pt,preprint]{aastex}%
\usepackage{epsf}
\usepackage{psfig}
\usepackage{graphics}

\usepackage{natbib}
\newcommand{\NH}{\mbox{${\rm N}_{\rm H}$}~} 

\begin{document}

\title{Suzaku Observations of Four Heavily Absorbed HMXBs}\label{Suzaku_HMXB}

\author{D.C.~Morris\altaffilmark{1,2}, R.K.~Smith\altaffilmark{2,3}, C.B.~Markwardt\altaffilmark{4,5}, R.F.~Mushotzky\altaffilmark{2}, J.~Tueller\altaffilmark{2}, T.R.~Kallman\altaffilmark{2}, K.S.~Dhuga\altaffilmark{1}}

\altaffiltext{1}{Center for Nuclear Studies, Department of Physics, The George Washington University, Washington, DC 20052, USA; {\it david.c.morris@nasa.gov}}
\altaffiltext{2}{NASA/Goddard Space Flight Center, Greenbelt, MD 20771, USA}
\altaffiltext{3}{Department of Physics and Astronomy, The Johns Hopkins University, Baltimore, MD. 21218}
\altaffiltext{4}{CRESST and Astroparticle Physics Laboratory, NASA/GSFC, Greenbelt, MD 20771, USA}
\altaffiltext{5}{Department of Astronomy, University of Maryland, College Park, MD 20742, USA}

\begin{abstract}
We report on Suzaku observations of four unidentified sources from the INTEGRAL and {\it Swift} BAT
Galactic plane surveys. All the sources have a large neutral hydrogen column density and are likely members of an
emerging class of heavily absorbed high mass X-ray binary (HMXB) first identified in INTEGRAL observations. 
Two of the sources in our
sample are approximately constant flux sources, one source shows periodic variation and one source exhibited
a short, bright X-ray outburst. The periodicity is transient, suggesting it is produced by a neutron star 
in an elliptical orbit around a stellar wind source. We analyze the flaring source in several segments to look for
spectral variation and discuss the implications of the findings for the nature of the source. We conclude
that all four sources in our sample can be identified with the emerging class of highly absorbed HMXBs, that one is 
a newly identified transient X-ray pulsar and that at least one is a newly identified supergiant fast X-ray transient (SFXT). 
\end{abstract}


\section{Introduction}

With its first observation, INTEGRAL \citep{Winkler03} began a collection of highly absorbed
sources, some of which may represent a new class of binary (see \cite{Kuulkers05} for a review; K05 hereafter). 
IGRJ16318-4848 was a bright X-ray source with extreme absorption that would later be shown to have 
bright X-ray emission lines. Since then, the INTEGRAL Galactic plane survey has
discovered several more similar sources, suggesting that these objects form a new class 
of X-ray binary, characterized by high absorption columns, slow spin periods and occasional bright X-ray 
outbursts. The first INTEGRAL catalog \citep{Bird04} notes 28 unidentified sources from the INTEGRAL 
Galactic plane survey, ten of which have been followed up at X-ray and optical wavelengths and have been 
shown to belong to this new source class. 

Nearly simultaneously to the discovery of this potentially new class of source by INTEGRAL, the {\it Swift} 
BAT Galactic plane survey began to uncover many new sources from its own survey. It was 
speculated, due to their absence from previous soft X-ray surveys, that some of these sources may be 
highly absorbed binaries, possibly members of the new class of INTEGRAL binaries.

Several objects identified in the first INTEGRAL catalog, as of Fall 2006, remained without significant 
follow-up in the X-ray or at other wavelengths. During the Suzaku cycle 1 call for proposals, we 
proposed to follow-up three INTEGRAL detected sources that were suspected to be members of the new class but
had yet to be studied in detail. We proposed, furthermore, to observe two newly identified sources from 
the BAT Galactic plane survey to determine whether they were also members of this new source class. 

In this paper, we detail the analysis of the Suzaku data collected on these sources, discuss the 
likelihood that these sources are similar to the originally discovered IGR sources and briefly discuss the 
possible nature of the new class. The paper is organized as follows: in \S2 we describe the 
observations and data analysis; in \S3 we present our results; in \S4 we discuss the 
similarity of each source and the propriety of calling each a member of the IGR source class and discuss 
the potential nature of these sources; in \S5 we summarize our results and conclusions.
 
\section{Observations and Data Reduction}

Observations of the five targets, three INTEGRAL sources and two additional sources from the BAT Galactic plane 
survey, were conducted between April 12 and October 31, 2006 (see Table 1). On four of the five targets, 
a single observation was collected while on one source two observations 
were collected, separated by $\sim$6 months. 
All sources were observed using the HXD aimpoint and with the XIS instruments in normal 
imaging mode. One of the sources (SWJ1010.1-5747) was found to be a symbiotic star and these observations
were published in \cite{Smith08}; this source will not be discussed further here. 

Data from the four other targets were reduced using the standard Suzaku processing software, xisrmfgen 
version 2007-05-14 and xissimarfgen version 2007-09-22 and the other standard analysis tools contained 
in HEADAS version 6.4. Point source processing was carried out as described in the Suzaku ABC 
Data Reduction Guide \footnotemark\footnotetext{http://heasarc.gsfc.nasa.gov/docs/suzaku/analysis/abc}.  
In processing the data from the HXD instrument, which is a non-focused instrument, it was discovered that 
the observation of IGRJ16465-4507 was contaminated by the nearby bright X-ray source GX340+0. 
As a result the HXD instrument data for this source have not been used. The HXD data of 
the other sources were checked for contamination by nearby bright sources but none was found. 

Data from the XIS0, XIS2 and XIS3 front illuminated instruments were combined together while data from 
the XIS1 back illuminated instrument were kept separate. Data from the combined front illuminated 
instruments, from the back illuminated instrument and from the HXD PIN (except in the case of 
IGRJ16465-4507) were then fit simultaneously in XSPEC (version 12.4.0). Data from the XIS instruments 
were binned to have a minimum of 50 events per bin. Data from the HXD instruments were binned to have a 
minimum of 100 events per bin. 

The Xronos timing analysis software version 5.21 was used to search for periodicity in the data. In 
cases where significant temporal structure was found in the data (either periodicity or transient 
outbursts), time-separated spectral analysis was done to search for spectral variation.

\begin{deluxetable}{cccccc}
\tablecaption{Observational Parameters}
\tabletypesize{\normalsize} \tablecolumns{6} \tablewidth{0pt}
\tablehead{ \colhead{Obsnum} & \colhead{Source} & \colhead{observation start} & \colhead{observation stop} & \colhead{XIS $\Delta$t} & \colhead{HXD PIN $\Delta$t}\\
\colhead{} & \colhead{} & \colhead{} & \colhead{} & \colhead{(s)} & \colhead{(s)}} \startdata
\hline
\hline
401052010 & IGRJ16465-4507 & 2006-09-09-09:12:56 & 2006-09-09-22:05:14 & 14536 & 24645\\
401053010 & SWJ2000.6+3210 & 2006-04-12-15:53:10 & 2006-04-12-21:56:04 & 12444 & 9877\\
401053020 & SWJ2000.6+3210 & 2006-10-31-00:29:37 & 2006-10-31-07:16:19 & 10146 & 11727\\
401054010 & IGRJ16493-4348 & 2006-10-05-21:10:30 & 2006-10-06-10:05:24 & 18975 & 20220\\
401055010 & SWJ1010.1-5747 & 2006-06-05-05:13:12 & 2006-06-05-18:25:25 & 19171 & 20000\\ 
401056010 & IGRJ16195-4945& 2006-09-20-20:25:12 & 2006-09-21-17:21:20 & 27908 & 42265\\
\hline
\enddata
\end{deluxetable}

\section{Results}

Two of the sources in our sample show only random temporal variability while two of the 
sources show important temporal and spectral variations that warrant further discussion. We begin this 
section with an overview of the characteristics of each source, considering the observation as a whole, 
and the global characteristics of the sample. We will then focus, in turn, on each of the two sources 
whose data are worthy of more detailed analysis.

\subsection{Global fitting}

In some cases, fitting a simple absorbed powerlaw model produces a fit with $\chi^2_{\nu}\sim$1, but in 
all cases a partial covering absorber model is strongly preferred (see Table 2). 
The partial covering model is invoked here for its ability to quantify interesting 
parameters from a wide range of geometries using a minimum of fit components. 
Assuming a dust halo subtending 4$\pi$ steradians, for example, the partial covering
model provides information about the scattering fraction and thus about the density of the
halo. Assuming, instead, a geometry similar to accretion disk corona (ADC) systems, the partial
covering model provides information about the vertical extent of the ADC and the scale height
of the neutral disk material. While the precision of the observations discussed in this
paper is too low to confidently distinguish between these different possible geometries, 
we adopt the use of the partial covering model here to lend our results to future 
discussion in this context.

The photon index ($\Gamma$)
in the partial covering model, ranges from 1.8 to 2.4, the partial covering fraction (PCF) ranges from 0.5 
to 0.8 and the \NH column ranges from $\sim$1$\times$10$^{23}$~cm$^{-2}$ to $\sim$1$\times$10$^{24}$~cm$^{-2}$. These $\Gamma$
are similar to those seen previously in highly absorbed HMXBs \citep{Kuulkers05}. In one case, the \NH 
column is slightly below 1$\times$10$^{23}$~cm$^{-2}$ (though within errors; 9$\times$10$^{22}$~cm$^{-2}$). 
We note that the HXD data for this 
target, IGRJ16465-4507, are 
contaminated by a field source and cannot be used. As a result, the high energy spectral slope
for this source is not well constrained as it is for the others, probably leading to an unusually
high value of the PCF compared to the other sources. The effect of the higher PCF
is to suppress the soft X-ray flux which helps to explain the lower \NH column in this source. 
If we assume a PCF and $\Gamma$ 
for this source similar to the other three sources in our sample (defined by the mean values of 
PCF=0.6 and $\Gamma$=2.15 respectively) we find \NH column values of 3.2$\times$10$^{22}$~cm$^{-2}$ and 1.0$\times$10$^{23}$~cm$^{-2}$ 
for the Galactic and local components respectively, similar to what is seen in the other three sources. 

\begin{deluxetable}{cccccccccc}
\tablecaption{All sources Partial Covering Absorber Fits\label{T7}}
\tabletypesize{\scriptsize} \tablecolumns{10} \tablewidth{0pt}
\tablehead{\colhead{Source} & \colhead{\NH} & \colhead{\NH$_{\rm{part}}$} & \colhead{PCF} & \colhead{$\Gamma$} & \colhead{Flux} & \colhead{Fe EW} & \colhead{$\chi^2_{\nu_{PC}}$} & \colhead{$\chi^2_{\nu_{PL}}$} & \colhead{dof}\\
\colhead{} & \colhead{(10$^{22}$cm$^{-2}$)\tablenotemark{a}} & \colhead{(10$^{22}$cm$^{-2}$)\tablenotemark{b}} & \colhead{\tablenotemark{c}} & \colhead{\tablenotemark{d}} & \colhead{(10$^{-12}\frac{\rm{ergs}}{\rm{cm^{2}~s}}$)\tablenotemark{e}} & \colhead{eV} & \colhead{\tablenotemark{f}} & \colhead{\tablenotemark{g}} & \colhead{\tablenotemark{h}}} \startdata
\hline
\hline
IGRJ16465-4507 & 2.0$_{-0.7}^{+0.7}$ & 7.3$_{-2.1}^{+2.8}$ & 0.82$_{-0.09}^{+0.06}$ & 2.19$_{-0.31}^{+0.35}$ & 8.95$_{-3.1}^{+0.2}$ & $<$135 & 0.859 & 0.994 & 270\\
IGRJ16493-4348 & 8.6$_{-1.0}^{+0.9}$ & 26$_{-7.9}^{+9.4}$ & 0.62$_{-0.07}^{+0.06}$ & 2.37$_{-0.17}^{+0.18}$ & 13.5$_{-2.0}^{+0.3}$ & $<$84 & 0.902 & 1.09 & 389\\ 
IGRJ16195-4945 & 11$_{-1}^{+1}$ & 78$_{-17}^{+17}$ & 0.53$_{-0.10}^{+0.09}$ & 1.80$_{-0.13}^{+0.14}$ & 16.1$_{-2.5}^{+0.2}$ & $<$43 & 0.934 & 0.999 & 614\\
SWJ2000.6+3210(1) & 2.3$_{-0.2}^{+0.2}$ & 9.3$_{-1.0}^{+1.2}$ & 0.68$_{-0.02}^{+0.03}$ & 2.21$_{-0.09}^{+0.08}$ & 32.0$_{-0.9}^{+0.5}$ & 51$^{+34}_{-37}$ & 0.918 & 1.58 & 417\\
SWJ2000.6+3210(2) & 2.1$_{-0.2}^{+0.2}$ & 8.2$_{-1.0}^{+1.1}$ & 0.70$_{-0.03}^{+0.02}$ & 2.01$_{-0.05}^{+0.05}$ & 55.0$_{-0.6}^{+0.8}$ & 71$^{+32}_{-29}$ & 0.897 & 1.66 & 725\\

\hline
\hline
\enddata
\tablenotetext{a}{\begin{footnotesize}Fully covered neutral hydrogen column\end{footnotesize}}
\tablenotetext{b}{\begin{footnotesize}Partially covered neutral hydrogen column\end{footnotesize}}
\tablenotetext{c}{\begin{footnotesize}Partial covering fraction\end{footnotesize}}
\tablenotetext{d}{\begin{footnotesize}Photon index\end{footnotesize}}
\tablenotetext{e}{\begin{footnotesize}Fit range set to 0.2keV to 10 keV\end{footnotesize}}
\tablenotetext{f}{\begin{footnotesize}Reduced $\chi^2$ of partial covering fit\end{footnotesize}}
\tablenotetext{g}{\begin{footnotesize}Reduced $\chi^2$ of simple absorbed powerlaw fit\end{footnotesize}}
\tablenotetext{h}{\begin{footnotesize}degrees of freedom in partial covering fit\end{footnotesize}}
\end{deluxetable}

All sources show variability on timescales of hundreds of seconds. Furthermore, IGRJ16195-4945 
shows a bright outburst lasting $\sim$5000~s and SWJ2000.6+3210 shows periodic variations. We will return 
to discuss these two objects in greater detail shortly. Overall flux levels range from 1$\times$10$^{-11}$ 
to 5$\times$10$^{-11}$ ergs~cm$^{-2}$~s$^{-1}$ (absorbed) in the 0.2-10 keV energy band. 

As a check of consistency with previous work, we have compared the results of absorbed 
cutoff powerlaw fits of the three sources in our sample that were also fit to cutoff powerlaws by 
K05. Our best fit results for IGRJ16195-4945 match well. Our best fit results for 
IGRJ16493-4348 do not match those of K05, but if we fix $\Gamma$ to be similar to that 
found by K05, we find a column density similar to theirs and can additionally note a cutoff 
energy of 17.5 keV. Our best fit results of IGRJ16465-4507 do not match that of K05. Even 
after fixing $\Gamma$ and the cutoff energy to match that of K05, the \NH column that we 
find is still more than an order of magnitude lower than that found by K05. As
noted earlier, the observation of IGRJ16465-4507 suffered from high energy contamination, leaving us with 
data only form 0.2-10.0 keV. While this may account for some of the discrepancy between our results and 
K05, we also note that we see no evidence in our data of the $\sim$4 minute period noted by 
K05. This suggests that our data may be taken during the apoastron phase of an elliptical binary orbit when 
the flux level, spectrum and absorption column have different values than they do 
near periastron and where the higher wind density is likely to produce periodic variations as are reported in the 
K05 observation. Thus, it is not surprising that our observations of this source appear 
different. In general, however, our overall results appear consistent with those of K05.

{\it Swift} BAT data from the forthcoming 22-month survey \citep{Tueller08} have been analyzed for each of the 
sources in our sample except for IGRJ16465-4507 which is contaminated by the same nearby source which hampers
the Suzaku HXD analysis. No evidence of outbursts on daylong timescales (similar to those seen in the 
Suzaku observation of IGRJ16915-4945, e.g.) are found. The BAT daily survey detection threshold 
is $\sim$5-10 mCrab \citep{Markwardt05},
however, so the non-detection of the $\sim$100-500 $\mu$Crab outbursts seen in
our Suzaku data is not surprising. 
We have, furthermore, examined the BAT data binned over longer timescales (2, 4, 8, 16, 32 and 64 days)
to improve the sensitivity of our variability search. 
While all sources show stochastic variability on these timescales, no evidence of periodicity is seen. 

For the 3 sources with uncontaminated BAT data, we have made simultaneous fits with the Suzaku and BAT
data. In the cases of IGRJ16493-4348 and SWJ2000.6+3210, the
BAT data are well fit by the same model derived from fitting the Suzaku data alone. This implies that the
snapshot captured in the Suzaku observations of these 2 sources is representative of the overall source
behavior and shows that the partial covering scenario is supported to energies as high as 200keV. In the
case of IGRJ16195-4945, the BAT data are not well fit to the same spectrum as the total Suzaku observation, nor to
any of the segmented spectra shown in Table 5. Since the total Suzaku observation does not match the
BAT data, we can infer that flares such as that seen in the Suzaku data do not dominate the flux from the 
source. Since
the BAT data are also not well fit by the spectrum seen in the Suzaku quiescent data (the BAT data fit a softer
spectrum than any segment of the Suzaku data and have a higher intensity than the quiescent segment but lower intensity than
the Post-Flare segment), we can also infer that the quiescent state is not best representative of this source. 
Since the post-flare state is also a poor fit to the BAT data, we infer that the source spends more time
in the quiescent state than is seen in the Suzaku observation, but that flares such as that seen in the Suzaku
data are somewhat common (since the quiescent state does not match the BAT data). We will discuss this in
further detail in \S4.

Spectra and lightcurves (in cases where the lightcurve shows noteworthy behavior) 
of each source are shown in Figures 1-5. In the spectral plots, black datapoints show Suzaku data while
green datapoints show the average 22-month survey data for comparison. We note that the spectral
fits detailed in Tables 2, 4 and 5 are derived using Suzaku data only.

\subsection{IGRJ16195-4945}

The lightcurve of IGRJ16195-4945 is marked by a short, bright flare, seen in both the XIS and HXD 
instruments. The flare lasts for $\sim$5000~s and reaches a peak flux level $\sim$10$\times$ brighter than the 
prior emission level. Such flaring behavior is characteristic of the subset of absorbed HMXBs known as 
Supergiant Fast X-ray Transients (SFXTs, \cite{Negueruela06, Smith06}). 
To better understand the nature of the flare and how it compares 
to the source during quiescence, we have separated the data as shown in Figure 3 and Table 3. 
The data segmentation separates the period prior to the flare (quiescence) from the flare itself (flare) 
and from the period after the giant flare has decayed to near the original level (post-flare). 
The definition of the 
quiescent period is clear due to an extended gap in the data (20ks) prior to the onset of the giant 
flare. The definition of the end of the flare is less clear, but we have chosen to define the end of the 
flare with the end of the orbit during which the flare decays below $\sim$2$\times$ the original level. 
The subsequent data to the end of the observation are defined as post-flare. This definition ensures that
the flare segment is dominated by flare emission even if 
some small amount of ''post-flare'' emission is included. It also ensures that the post-flare data are 
sufficient to produce an accurate spectrum. We have further subdivided the giant 
flare into the rising leg of the flare, the decaying leg of the flare (which are conveniently 
separated by an orbital gap) and a segment that encompasses all data from
the onset of the flare to the end of the observation (''on'').

\begin{table}[h]
\centering
\caption{IGRJ16195-4945 Segmentation - Data from IGRJ16195-4945 are segmented to separate the period 
prior to the flare (quiescence) from the flare itself (flare) and from the period after the giant 
flare has decayed to near the original level (post-flare).}
\begin{tabular}{c c c c}
\hline\hline
Segment & start & stop & duration\\
& (s) & (s) & (s)\\
\hline
quiescence & 0 & 12000 & 12000\\
total flare & 40068 & 46684 & 6616\\
flare onset & 40068 & 41769 & 1701\\
flare decay & 44669 & 46684 & 2015\\
post-flare & 49000 & 72000 & 23000\\
on & 40068 & 72000 & 31932\\
\hline
\end{tabular}
\label{table:Segs}
\end{table}

The results of absorbed powerlaw and partial covering model spectral fits to each of the six segments 
(quiescence, total flare, flare onset, flare decay, post-flare, on) are shown in Table 4. 
To limit the number of free parameters in the fit we have fixed the global \NH column density, 
the partial covering column and the PCF to the values that they have when fitting the complete 
observation, allowing only $\Gamma$ to vary. We will return momentarily to the 
possibility that the spectral variation is due to these other parameters. 

When allowing only $\Gamma$ to vary, we find that the ''on'' segment is 
well fit by a model with $\Gamma\sim$1.8. The quiescent data, in 
contrast, are fit by a much softer $\Gamma\sim2.5$. No significant Fe 
lines are seen, although upper limits are given in Table 4. 
It is interesting to note that $\Gamma$ remains nearly unchanged (perhaps even 
becoming harder) as the flare subsides and the ''post-flare'' phase begins. Considering the modest flux 
level of the ''post-flare'' phase, one would expect that as the flare decays the spectrum would soften 
as the quiescent component once again becomes comparable to the flaring component. That this is 
not seen suggests 
that, during the ''post-flare'' phase, the emission component that was seen during the quiescent phase is 
not merely being overwhelmed by the component responsible for the flare emission, but rather that the 
quiescent component is absent altogether. 

\begin{deluxetable}{cccccccccc}
\tablecaption{J16195-4945 Time Resolved Partial Covering Absorber fits; Fixed \NH and covering fraction\label{T7}}
\tabletypesize{\normalsize} \tablecolumns{10} \tablewidth{0pt}
\tablehead{\colhead{Seg\#}&  \colhead{Source} & \colhead{\NH} & \colhead{\NH$_{\rm{part}}$} & \colhead{PCF} & \colhead{$\Gamma$} & \colhead{Flux} & \colhead{$\chi^2$} & \colhead{dof} & \colhead{Fe EW}\\
\colhead{} & \colhead{} & \colhead{(10$^{22}$cm$^{-2}$)} & \colhead{(10$^{22}$cm$^{-2}$)} & \colhead{} & \colhead{} & \colhead{(10$^{-12}\frac{\rm{ergs}}{\rm{cm^{2}~s}}$)\tablenotemark{a}} & \colhead{} & \colhead{} & \colhead{(eV)}} \startdata
\hline
\hline
1-6 & Total & 11 & 76 & 0.55 & 1.81$_{-0.03}^{+0.03}$ & 16.1$_{-1.0}^{+0.1}$ & 565.0 & 591 & $<$43\\
1 & quiescent & 11 & 76 & 0.55 & 2.46$_{-0.44}^{+0.48}$ & 2.68$_{-1.13}^{+0.25}$ & 64.5 & 88 & $<$320\\
2 & total flare & 11 & 76 & 0.55 & 1.78$_{-0.03}^{+0.03}$ & 48.0$_{-0.8}^{+0.7}$ & 290.7 & 340 & $<$19\\
3 & flare onset & 11 & 76 & 0.55 & 1.81$_{-0.12}^{+0.10}$ & 58.5$_{-1.9}^{+2.0}$ & 71.3 & 123 & $<$65\\ 
4 & flare decay & 11 & 76 & 0.55 & 1.82$_{-0.08}^{+0.08}$ & 45.8$_{-1.4}^{+1.5}$ & 99.4 & 122 & $<$37\\
5 & post-flare & 11 & 76 & 0.55 & 1.73$_{-0.06}^{+0.06}$ & 9.68$_{-0.25}^{+0.34}$ & 242.2 & 269 & $<$81\\
6 & on & 11 & 76 & 0.55 & 1.75$_{-0.03}^{+0.03}$ & 20.6$_{-0.3}^{+0.3}$ & 508.5 & 554 & $<$28\\
\hline
\hline
\enddata
\tablenotetext{a}{\begin{footnotesize}Fit range set to 0.2keV to 10 keV\end{footnotesize}}
\end{deluxetable}

If we leave all components of the partial covering model free to vary we find the results shown in Table 5. 
Here we see the quiescent phase described by a high PCF, moderately hard powerlaw 
and low column density. In contrast, the flare shows a low PCF and softer powerlaw with 
much higher column density. Finally, the ''post-flare'' phase shows a PCF 
between the other two, the softest $\Gamma$ and the highest column density of all the segments. 
This suggests that the flare signals the onset of emission from a region of much
greater local column density, perhaps from interaction with a disk. We will return to 
discuss this possibility in \S4. 

Previous observations of this source by both ASCA \citep{Sidoli05} and INTEGRAL \citep{Sguera06} have also shown
evidence of outbursts lasting $\sim$1-2 hours. Translated into a common
energy range the flux of these previous outbursts is lower than that seen in our Suzaku observations 
by a factor of a few, but differences in the measured spectral parameters and the lower signal to noise of these
earlier observations limit the precision of the comparison. Nevertheless, the similarity of the outbursts
in these three observations seems suggestive either of a characteristic timescale of the accretion flow of this 
source (perhaps indicative of an associated clump size in the flow) or of a periodic interaction of the neutron star
with an over-dense region in the wind, possibly a disk. We note that a short ($<$5ks) Chandra 
observation has been made of this source but no variability has previously been reported, though 
irregularities in the data complicate the analysis \citep{Tomsick06}.

\begin{deluxetable}{cccccccccc}
\tablecaption{J16195-4945 Time Resolved Partial Covering Absorber fits\label{T7}}
\tabletypesize{\normalsize} \tablecolumns{10} \tablewidth{0pt}
\tablehead{\colhead{Seg\#}  & \colhead{Source} & \colhead{\NH} & \colhead{\NH$_{\rm{part}}$} & \colhead{PCF} & \colhead{$\Gamma$} & \colhead{Flux} & \colhead{$\chi^2$} & \colhead{dof} & \colhead{Fe EW}\\
\colhead{} & \colhead{} & \colhead{(10$^{22}$cm$^{-2}$)} & \colhead{(10$^{22}$cm$^{-2}$)} & \colhead{} & \colhead{} & \colhead{(10$^{-12}\frac{\rm{ergs}}{\rm{cm^{2}~s}}$)\tablenotemark{a}} & \colhead{} & \colhead{} & \colhead{(eV)}} \startdata
\hline
\hline
1-6 & Total & 11$_{-1}^{+1}$ & 77$_{-18}^{+17}$ & 0.55$_{-0.10}^{+0.09}$ & 1.82$_{-0.13}^{+0.14}$ & 16.1$_{-2.4}^{+0.1}$ & 573.8 & 615 & $<$43\\
1 & quiescent & 3.5$_{-3.0}^{+6.0}$ & $<$10 & 0.95\tablenotemark{b} & 1.58$_{-0.5}^{+1.1}$ & 2.85$_{-2.85}^{+0.21}$ & 64.5 & 85 & $<$356\\
2 & total flare & 11$_{-1}^{+1}$ & 72$_{-18}^{+17}$ & 0.57$_{-0.10}^{+0.08}$ & 1.82$_{-0.15}^{+0.15}$ & 48.2$_{-8.2}^{+0.7}$ & 289.0 & 337 & $<$15\\
3 & flare onset & 10$_{-2}^{+2}$ & 48$_{-24}^{+23}$ & 0.52$_{-0.22}^{+0.16}$ & 1.73$_{-0.26}^{+0.30}$ & 59.4$_{-27.1}^{+2.4}$ & 69.0 & 120 & $<$49\\ 
4 & flare decay & 11$_{-2}^{+2}$ & 65$_{-28}^{+36}$ & 0.58$_{-0.21}^{+0.14}$ & 1.89$_{-0.30}^{+0.33}$ & 46.4$_{-28.6}^{+1.4}$ & 95.7 & 119 & $<$25\\
5 & post-flare & 12.4$_{-1.7}^{+1.8}$ & 121$_{-31}^{+30}$ & 0.75$_{-0.18}^{+0.11}$ & 2.07$_{-0.32}^{+0.33}$ & 9.73$_{-2.54}^{+2.18}$ & 237.0 & 266 & $<$68\\
6 & on & 11$_{-1}^{+1}$ & 83$_{-16}^{+16}$ & 0.55$_{-0.10}^{+0.08}$ & 1.75$_{-0.14}^{+0.14}$ & 20.4$_{-2.7}^{+0.3}$ & 507.2 & 551 & $<$43\\
\hline
\hline
\enddata
\tablenotetext{a}{\begin{footnotesize}Fit range set to 0.2keV to 10 keV\end{footnotesize}}
\tablenotetext{b}{\begin{footnotesize}PCF is poorly constrained in the quiescent fit due to the low number of counts\end{footnotesize}}
\end{deluxetable}

\subsection{SW2000.6+3210}

Data were collected on SWJ2000.6+3210 during two epochs, separated by six months (see Figures 4-5). 
We have analyzed each of these observations separately. Neither observation is 
adequately fit by a simple absorbed powerlaw (see Table 2). Both observations are well fit by a 
partial covering model and the parameters of both fits are quite similar. The 
later observation has a slightly harder $\Gamma$ and significantly larger flux. Both observations 
show evidence of a weak Fe fluorescence line at 6.4 keV with apparently constant equivalent width. 

Interestingly, while the first observation shows only random variability, the second observation shows 
regular variations with a period of 1056~s. We interpret this as the spin period of the 
neutron star companion in a binary, similar to long periodic variations noted in several other
sources \citep{Kuulkers05}. There is the suggestion of secular variation beneath the 1056~s period, 
but the data are insufficient to confidently determine any further periodic components.

This may suggest that the system is a HMXB with an elliptical orbit. During the first observation, the compact object is far from the donor star where the wind density is low. In this case, 
the X-ray flux is likely to be approximately constant, possibly associated with an accretion 
disk or corona around the compact object. During the second observation, the compact object is 
nearer to the donor star where the stellar wind is more dense, channeling more material onto 
the neutron star and producing the observed periodicity through accretion.

\section{Discussion}

All of the sources in our sample are generally consistent with being members of the new INTEGRAL highly 
absorbed binary source class. The most likely candidate among our sources to be a member of this new 
class is SWJ2000.6-3210, which displays moderately high \NH ($\sim$1$\times$10$^{23}$~cm$^{-2}$), $\Gamma$ and
cutoff energy similar to other members of the class and periodic variations with a 
period of 1056~s. 

IGRJ16195-4945 shows a high absorbing column and a brief but bright 
X-ray outburst, identifying it as a new SFXT. We note, however, that \NH variations are not strictly required
during the flare.  
If N$_{\rm{H}}$ variations are not invoked, the outbursts can be
associated with variable accretion and the duration of the flare may be indicative of the clump size in the companion
star wind. 

A recent alternative model \citep{Sidoli07} suggests that SFXT outbursts such as this
are due to periodic interaction with an equatorial disk wind from the donor star.  A dramatic increase in \NH 
column is seen when all parameters are left free to 
vary during fitting, supporting this scenario. Sidoli et al. suggest that such outbursts are actually
longer lived than the previously generally observed duration of hours, lasting instead for several days 
\citep{Romano07}.
Since our data coverage ends $\sim$30~ks after the flare onset, we cannot rule out that an elevated level of
emission continues for several days after the initial flare, though the brightest emission appears to last for
only $\sim$1 hour in our data. 

If we assume the equatorial disk wind model to be applicable to this source, however, we can estimate the orbital
period of the binary as follows. We assume that the three
broad states seen in the Suzaku observation (quiescent, post-flare and total flare) approximately represent the
full range of states of IGRJ16195-495 and thus we can write the following relationship between the flux of the 3 Suzaku states
and the orbit-averaged flux level seen by BAT:

\begin{equation}
a\times~a_{f}+b\times~b_{f}+c\times~c_{f}=d
\end{equation}

where a is the flux level of the Suzaku quiescent state observation and a$_{f}$ is the fraction of the orbit 
spent in this state, b is the flux level of the Suzaku post-flare state observation and b$_{f}$ is the fraction 
of the orbit spent in this state, c is the flux level of the Suzaku total flare state observation and c$_{f}$ is the
fraction of the orbit spent in this state and d is the (orbit averaged) flux of the BAT observation.
If we now make the assumption that the bright flare seen in the Suzaku data (the total flare state) represents 
the entry of the compact object into an equatorial disk wind from the donor star,
we must interpret the subsequent post-flare state as the longer lived period of moderate flaring behavior that
accompanies the remainder of the journey of the compact object through the disk wind. According to the \cite{Romano07}
observation of SwiftJ11215-5952, this longer lived period of activity can last for up to $\sim$5 days. Assuming a
similar duration for this phase in IGRJ16195-4945, we have b$_{f}=$4.3e5/P and c$_{f}=$5e3/P and a$_{f}=(1-\frac{4.35e5}{P})$
where P is the period of the binary in seconds. Thus equation (1) becomes:

\begin{equation}
P=\frac{-a\times4.35e5+b\times4.3e5+c\times5e3}{d-a}
\end{equation}

To estimate the flux levels a, b, c and d, we fit each corresponding dataset using the average spectral fit to the
complete dataset (including the BAT data) and determined the associated flux level in the 15-150~keV band. We find
a=5.13e-12, b=8.61e-11, c=3.93e-10 and d=3.11e-11~ergs/cm$^2$/s. Plugging these values into equation (2) 
returns a predicted period P$\sim$16 days. This period is unusually short compared to other measured 
SFXT periods \citep{Heras07, Romano07}. Even if we alter our calculation to interpret the on
state (see Tables 4-5) as the measure of the flux during the passage through the disk wind rather
than weighting it heavily to the post-flare state as shown above, the predicted period only increases by a factor
of $\sim$2, which is still well below the more typically reported SFXT orbital periods of $\gtrsim$200 days. An orbital
period of $\sim$16 days, however, is consistent with previous measurements of orbital periods associated
with highly absorbed IGR sources \citep{Corbet04, Corbet05}. 
We are pursuing further observations to followup this predicted orbital
period. If the predicted period is shown to be accurate, IGRJ16195-4945 may represent an important cross-over source between
the known SFXTs with orbital periods of hundreds of days with periodic flaring behavior and the new class of 
highly absorbed INTEGRAL sources, which have much shorter orbital periods but are without clear observations of periodic 
flaring behavior. Such a cross-over object would suggest a common mechanism at work for the emission seen from 
HMXBs of widely varying orbital parameters. It would also suggest a useful method of refining 
the period search for such sources in instances where a single flare has been observed but large amounts of
observatory time are not available for a dedicated follow-up monitoring campaign.
A long-term monitoring campaign of this source would be very useful to determine
the true duration of the associated outbursts.
 
IGRJ16493-4843 is a variable source with 
a high column density and non-periodic variations, also potentially qualifying it with 
this new source class. Finally, IGRJ16465-4507 presents the 
fewest similarities to previous observations of the class, but part of the discrepancy may be due to the 
narrow energy window in which our observations are made for this source. We note that we do not 
see periodic variability in this source as previously reported by others \citep{Lutovinov05, Kudryavtsev06}.

One of the interesting results of this work is that two of the four sources in 
our sample (IGRJ16465-4507 and SwiftJ2000.6+3210) appear to be transient pulsars. 
It would be very interesting 
to know if this is a common characteristic of all sources in this new class. If so, it would imply that
elliptical orbits are common in these sources. Since it has been shown, however, that binaries will 
circularize rapidly (10$^4$-10$^6$ years) through tidal dissipation following the
supernova detonation of the more massive star \citep{Savonije83, Lecar76}, the occurrence of highly elliptical orbits, 
if found to be common in these sources, will require explanation. As a rough estimate of the fraction of sources 
that may be expected to be in highly elliptical orbits, 
we may proceed from work showing that the companion source in several of these systems is a supergiant O or B star
\citep{Nespoli08b, Nespoli08a, Halpern06, Tomsick06, Negueruela05}. This implies that the HMXB
phase of the lifetime of the system (the time after the detonation of the more massive star and prior to the 
detonation of the less massive star) will be typically 5-10$\times$10$^6$ years. Therefore, at most 10-20\%
of the observed sources should be found in strongly elliptical orbits. Finding 
a significantly higher fraction of highly elliptical orbits would suggest another mechanism at work during the
HMXB evolution. Possible explanations for the lower circularization rate include 
generally wide orbital separation of the binaries, unusually large initial supernova kicks, and capture events or
other gravitational interactions that may amplify eccentricity of the system. 

It has been shown by \cite{Abt05} that the ellipticity of visual binaries is roughly a function of the mass
of the primary and the orbital period of the system, with tidal dissipation responsible for rapid circularization
of systems with orbital periods less than about 10 days. Since the same mechanism of tidal dissipation is thought to 
be responsible for the circularization of X-ray binary orbits, it seems reasonable to assume that a similar 
function applies to HMXBs. Measuring both the ellipticity of the orbit and the period for a sample of 
these systems through a monitoring campaign will allow us to test whether this hypothesis is true, and thus will
indicate whether tidal dissipation is the dominant effect in HMXB circularization. 

Monitoring would also 
determine orbital periods of each system, a necessary step toward determining the orbital characteristics 
of the system and ultimately the mass of the compact object, and would allow a more in-depth analysis of 
spectral variability as a function of orbital phase. Since the neutron stars in these binaries likely orbit only a
few stellar radii from the companion and are believed to have 
orbital periods on the order of ten days, daily or perhaps bi-weekly observations for about a month 
seem sufficient to accurately determine all of these characteristics. The {\it Swift} satellite offers the 
capability to perform such monitoring and, with its simultaneous X-ray and optical observations, would 
provide an ideal platform for future observations. Finally, monitoring on long timescales (covering a 
baseline of several years) will allow investigation of the spin-up/spin-down of the pulsar which 
can shed light on the magnetic field strength of the pulsar as well as the wind environment in which it is
embedded (see, e.g., \cite{Patel07}).

Three of the four sources in our sample are confidently associated with OB stars from optical spectroscopy
\citep{Negueruela05, Halpern06, Nespoli08a}. 
In the case of IGRJ16915-4945, a foreground dwarf star has thwarted attempts to identify 
the secondary. IGRJ16915-4945 has a history of X-ray outbursts, however, which are a trait of SFXTs that 
contain OB secondary stars \citep{Negueruela06}. Thus association with a massive star seems likely for this system as well.
Since the companion stars in these sources are most likely O or B stars, an interesting potential application 
of these data is to study the porosity or clump size of the absorbing material in the winds of 
massive stars. Though the periodic variability and extremely bright outbursts seen in some of the data
cannot be explained by variations in the column density along the line of the sight to the observer in classic HMXBs,
stochastic variations of lower contrast, which are also seen in the data, are a potential signature of 
variable obscuration. We have searched for variations in column density (assuming all other spectral parameters
remain constant as a simplifying approximation) during the aperiodic pulses seen in some of the data 
but do not find any to within the uncertainty of the data. We point out, however, that typical fluences
in the pulses analyzed are $<$1000 photons, leading to weak constraints on the column density. While some
of these sources have prior observations with XMM-Newton, whose greater effective area will reduce
the uncertainty in these measurements, the exposures are extremely short (3-4 ks) and thus are not able to 
constrain column density variations on timescales of 5-10 ks as seen in these data. 
Longer observations of these sources with XMM-Newton would be useful in this regard and would offer the
additional benefit of high resolution soft X-ray spectroscopy to probe the ionization
state of the wind through metal line strengths. \cite{Walter07} have investigated
the size of clumps in the accreting wind of SFXTs and determined typical masses of 10$^{22}$-10$^{23}$~g and mass loss
rates of 10$^{-5}$-10$^{-6}$ M$_\odot$/yr from analysis of INTEGRAL data. 
INTEGRAL data do not constrain N$_{\rm{H}}$, however, and thus
cannot distinguish between variability due to changing column density levels and that due to changing accretion
levels. XMM-Newton observations at 0.2-10.0~keV will be able to distinguish between these two variability mechanisms. 

The limits on the Fe line equivalent widths (EWs) found in these sources are low for sources of such high N$_{\rm{H}}$ column density
\citep{Makishima86}. Such low Fe line EWs imply either an intrinsically low Fe abundance relative to N$_{\rm{H}}$
(as has been seen in extragalactic sources, e.g., Centaurus A \citep{Markowitz07}) or that the Fe fluorescence
covering fraction is less than 1. If the high N$_{\rm{H}}$ column density absorber is local to the compact object,
as has been suggested through comparisons between X-ray and optical absorption measures (e.g., \cite{Revnivtsev03, Walter03, Chaty04}),
and if the Fe fluorescence originates in a shell structure surrounding the compact object, 
it is difficult to envision a geometry in which the coverage would be less than 4$\pi$. The Fe emission
may originate in the inner region of the accretion disk, however, in which case a lower covering
fraction may be reasonable. Furthermore, there are reasons to believe that the assumptions inherent in
relating X-ray and optical absorption measures may not be valid for all sources \citep{Maiolino01}, and thus the disagreement
between X-ray and optical absorption measures may not necessarily imply that the high column absorber is
local to the compact object. Alternatively, such low Fe EWs might also be considered as evidence of the presence of ADC
emission. The presence of an ADC seems unlikely, however, given the lack
of other emission lines from 1-5 keV that are usually produced in ADCs. Still other explanations
for low Fe EWs are similar to those discussed by \cite{Markowitz07} in reference to Cen A and include
the Fe fluorescence source being displaced from the X-ray continuum source or
the presence of an attenuating obstruction between the X-ray source and the site of the Fe fluorescence. While our
data argue against (but do not rule out) the low covering fraction scenario, the intrinsically low Fe EWs
scenario requires non-trivial assumptions about the geometry of the system. Further observations, particularly
phase-resolved spectroscopy, will be very helpful in disentangling these many potential explanations and are
encouraged. 

\section{Summary and Conclusions}

Our primary goal in this study is to determine whether these four sources exhibit similar characteristics to 
the emerging new class of highly absorbed IGR sgHMXB or SFXT sources. 
The defining characteristics of this new class are i) high absorption column ($>$1$\times$10$^{23}$~cm$^{-2}$), ii) 
periodicity on timescales of a few to $\sim$100 minutes, generally 
interpreted as a neutron star spin period, iii) periodicity on timescales of $\sim$10 days, generally 
interpreted as a binary orbital period and iv) occasionally strong X-ray emission lines. 

\noindent {\bf i)} In all four sources the total absorbing column 
is $\gtrsim$1$\times$10$^{23}$~cm$^{-2}$. Generally, 
we do not see significant changes in the absorption over the duration of the observations. While this is 
not surprising during a single day-long observation, which probably samples a small segment of 
the orbital phase, it is somewhat surprising that two observations of SWJ20006.+3210, separated by more 
than six months (and furthermore apparently at different parts of the orbital phase) also show no 
significant difference in \NH. A dramatic increase in absorbing column is seen in IGR16195-4945 for a 
duration of $\sim$30~ks, during a bright outburst (when all parameters are left free to vary). Since this dramatic increase in \NH is 
associated with a dramatic increase in flux, one might interpret this as an 
increase in emission due to interaction of the neutron star with a density enhancement. Such a scenario might be 
expected due to the passage of the neutron star through a thick disk associated with the donor star.
Since an interpretation in which \NH is approximately constant and only $\Gamma$
varies also produces adequate spectral fits, an alternative model in which the increased emission is due to
variable accretion is also possible.

\noindent {\bf ii)} In SWJ2000.6+3210, we 
see a period of 1056~s which is only observed during one of two observations of the 
source. This is a newly identified transient X-ray pulsar. Given the low 
magnetic field implied by the relatively slow rotation period of 1056~s, it is not surprising that the 
same mechanism that produces periodic observations near periastron is too weak to produce an observable 
period near apoastron where the wind density will be lower. 

\noindent {\bf iii)} Due to the short nature of the observations (generally $\sim$1 day), we do not expect to 
directly measure orbital periods, previously reported to be on the order of ten days in other sgHMXBs and, indeed, 
we do not. During 
observations of the one source for which we have well separated observations we also do not see evidence of 
an orbital period. We have, however, used our observations to calculate a predicted orbital period for 
IGRJ16195-4945. Assuming that IGRJ16195-4945 is a SFXT and that the bright flare we see signals the 
interaction of the compact object with an equatorial wind of the donor star, we have combined our Suzaku
observations with long-term {\it Swift}-BAT observations to predict the orbital period for this source as P$\sim$16 days.
This is unusually short compared to previously measured SFXT orbital periods, but is consistent with orbital periods
previously measured for the highly absorbed IGR sources. More consistent monitoring, perhaps using the {\it Swift} satellite, 
would be very useful in confirming or refuting this predicted period and in characterizing the orbital periods
of these highly absorbed binaries in general. 

\noindent {\bf iv)} We find only weak evidence of Fe fluorescence emission in one source, and only upper limits 
to Fe lines in the other three sources. This is similar to the behavior reported in the original INTEGRAL highly 
absorbed X-ray binary sources 
\citep{Kuulkers05} in which only 1 of 10 sources showed strong emission lines while the others showed only 
weak lines or upper limits. The Fe line measurements that we report here 
are either the first for the source in the literature (IGRJ16465-4507, IGRJ16493-4348, SwiftJ2000.6+3210) or several times
more restrictive than previous measurements (IGRJ16195-4945). Moreover, our measurements are similarly or more restrictive
than those reported on similar sources in the literature (see, e.g., \cite{Sidoli05, Patel04, Rodriguez03}).

While none of these sources is ruled out from the IGR class based on these observations, further 
observations would be helpful in refining their nature. Of particular value would be periodic 
monitoring of all four systems.

\acknowledgements
We are grateful to Hans Krimm and Gerry Skinner for providing access to and help in analyzing the BAT survey data and to Fotis Gavril for help in the timing analysis studies. We also thank Maurice Leutenegger for useful discussion and comments. D.~C.~M. acknowledges support from the Center for Nuclear Studies (CNS) through the Research Enhancement Fund at the George Washington University and from NASA grant NNH05ZDA001N-SUZ/11132.

\begin{figure}
\figurenum{1} \epsscale{0.8} \rotatebox{270}{\plotone{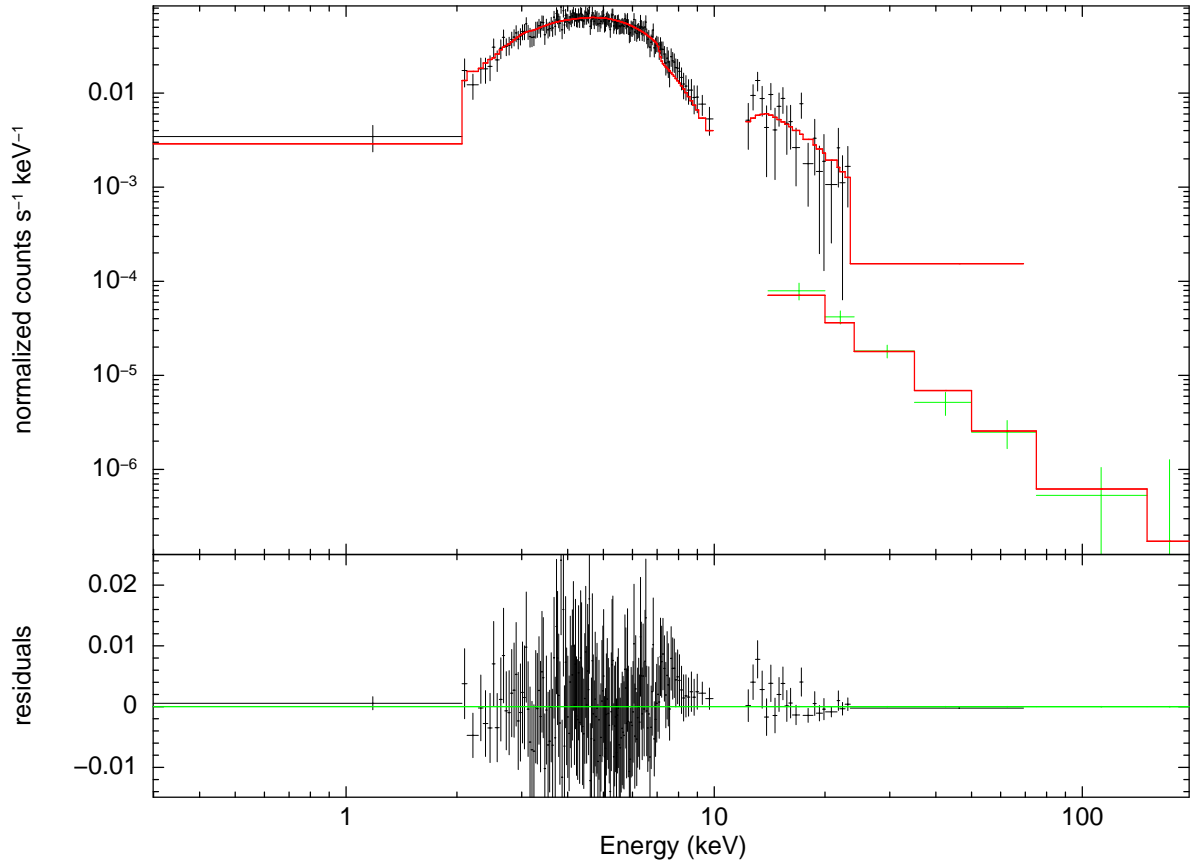}}
\caption[IGRJ16493-4348 spectrum in the 0.2-150.0 keV energy range. Low energy data are from the XIS0,XIS1,XIS2 and XIS3 instruments and high energy data from the HXD-PIN and {\it Swift}-BAT. Suzaku data are shown in black while BAT data are shown in green.]{\label{figure1}IGRJ16493-4348 spectrum in the 0.2-150.0 keV energy range. Low energy data are from the XIS0,XIS1,XIS2 and XIS3 instruments and high energy data from the HXD-PIN and {\it Swift}-BAT. Suzaku data are shown in black while BAT data are shown in green.}
\label{fig1}
\end{figure}

\begin{figure}
\figurenum{2} \epsscale{0.8} \rotatebox{270}{\plotone{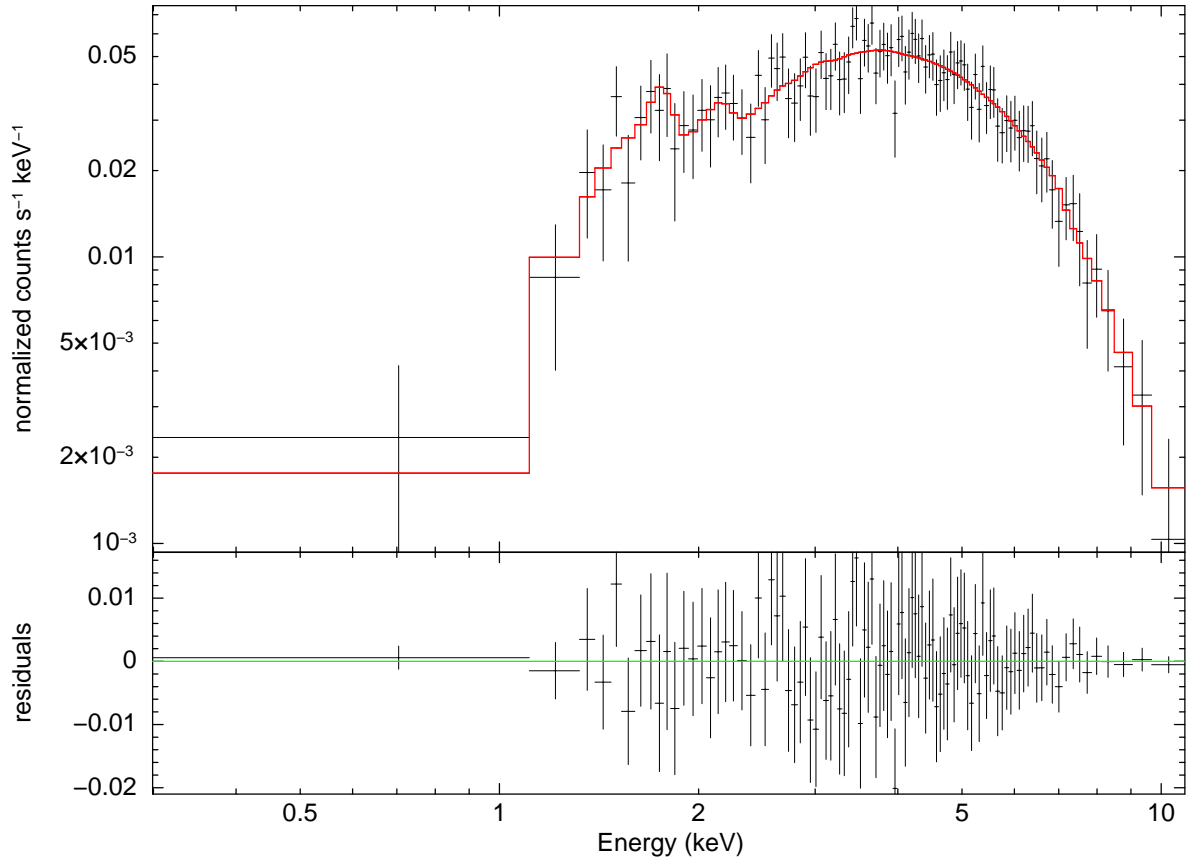}}
\caption[IGRJ16465-4507 spectrum in the 0.2-10.0 keV energy range. Low energy data are from the XIS0,XIS1,XIS2 and XIS3 instruments. High energy data from the HXD-PIN are contaminated by a bright serendipitous source and are therefore not included in the analysis.]{\label{figure2}IGRJ16465-4507 spectrum in the 0.2-10.0 keV energy range. Low energy data are from the XIS0,XIS1,XIS2 and XIS3 instruments. High energy data from the HXD-PIN are contaminated by a bright serendipitous source and are therefore not included in the analysis.}
\label{fig2}
\end{figure}

\begin{figure}[h]
\figurenum{3} \epsscale{0.6} \rotatebox{270}{\plotone{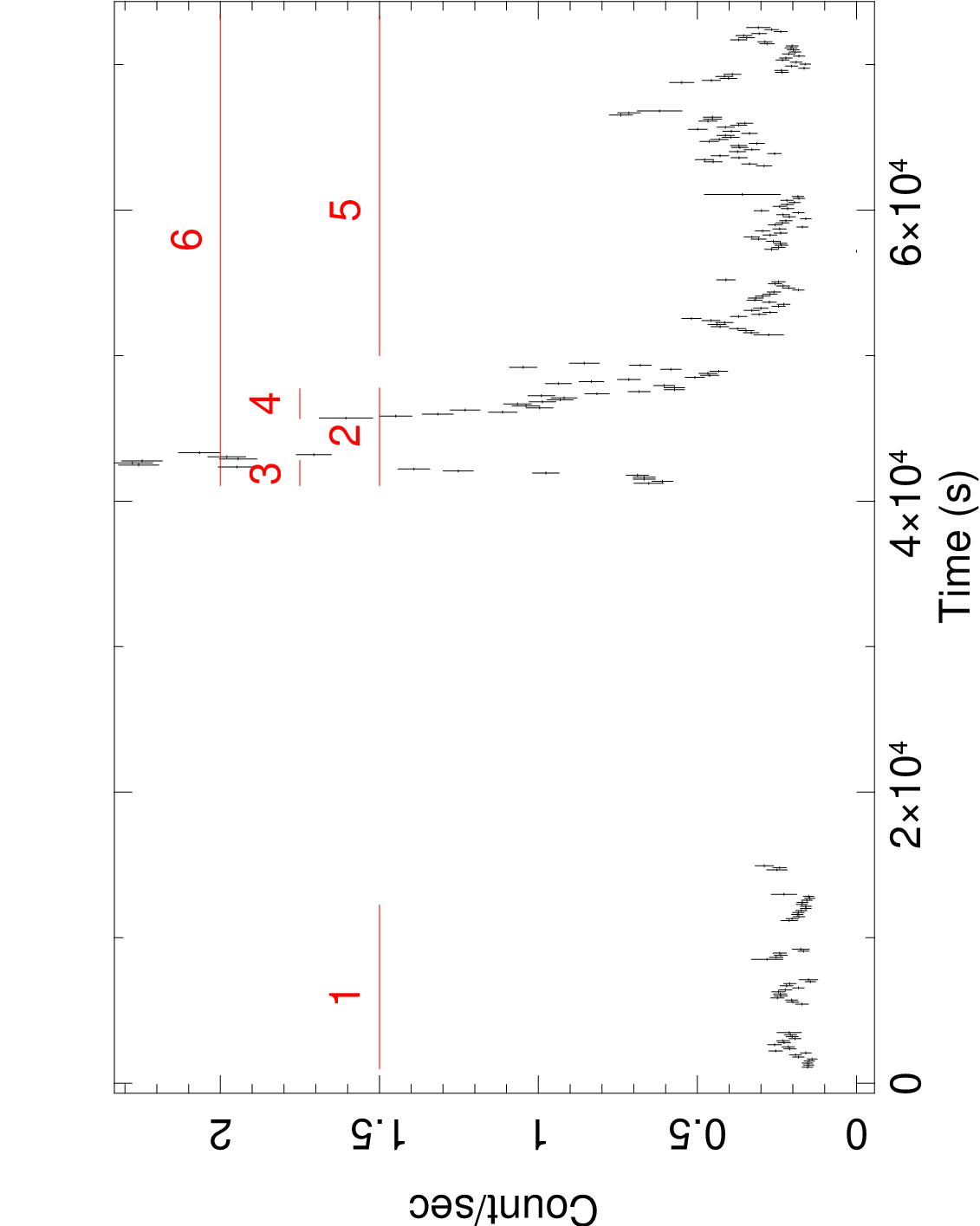}}
\epsscale{0.6} \rotatebox{270}{\plotone{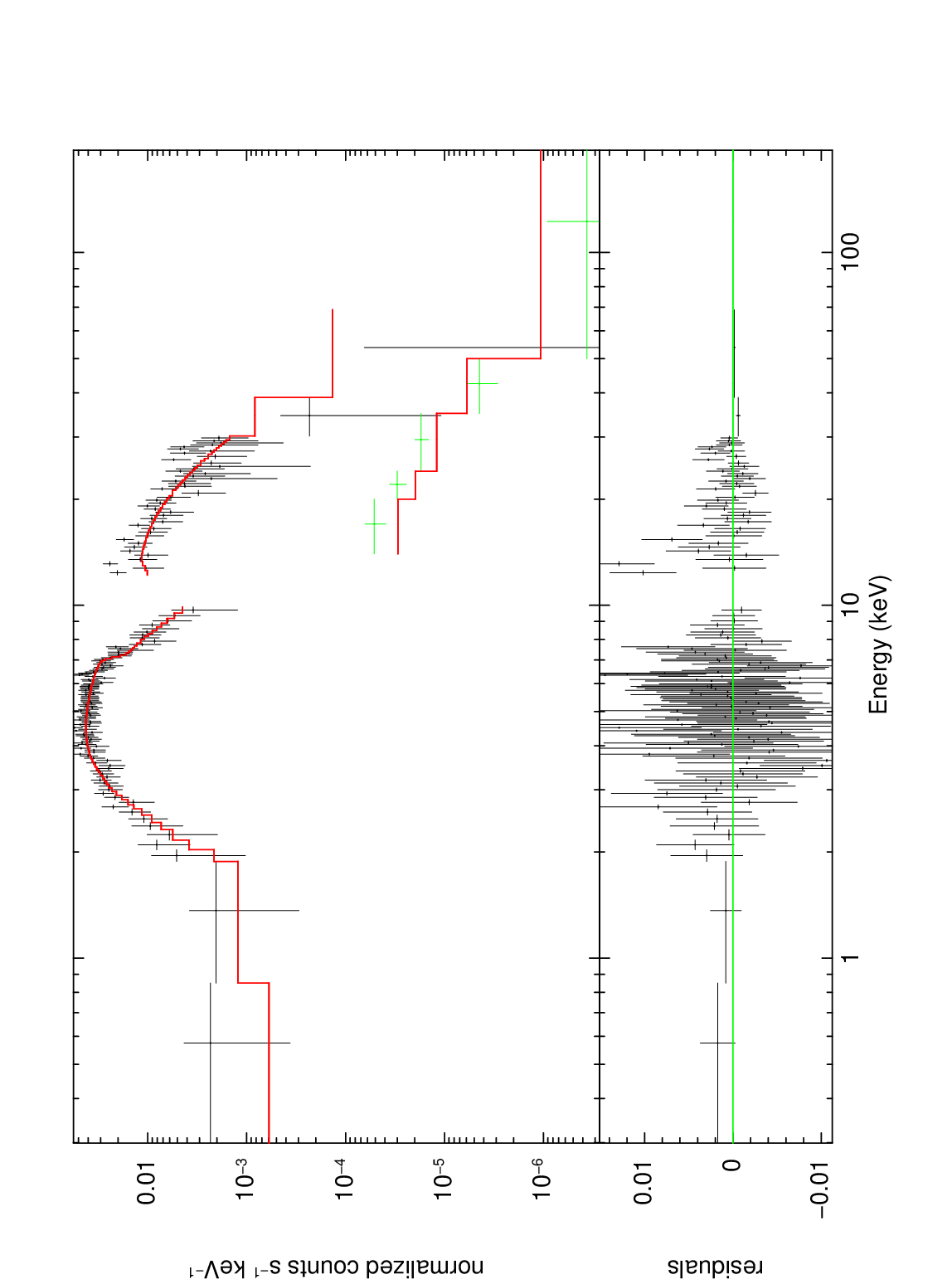}}
\caption[IGRJ16915-4945 Lightcurve (Top) and total spectrum in the 0.2-150.0 keV energy range (Bottom). Suzaku data are shown in black while BAT data are shown in green. Data are extracted and analyzed over the entire dataset but also in each of the data subsets shown in the figure by numbered red lines. The numbered segments correspond to separated fits as shown in Tables 4-5.]{\label{figure1}IGRJ16915-4945 Lightcurve (Top) and total spectrum in the 0.2-150.0 keV energy range (Bottom). Suzaku data are shown in black while BAT data are shown in green. Data are extracted and analyzed over the entire dataset but also in each of the data subsets shown in the figure by numbered red lines. The numbered segments correspond to separated fits as shown in Tables 4-5.}
\label{fig3}
\end{figure}

\begin{figure}
\figurenum{4} \epsscale{0.6} \rotatebox{270}{\plotone{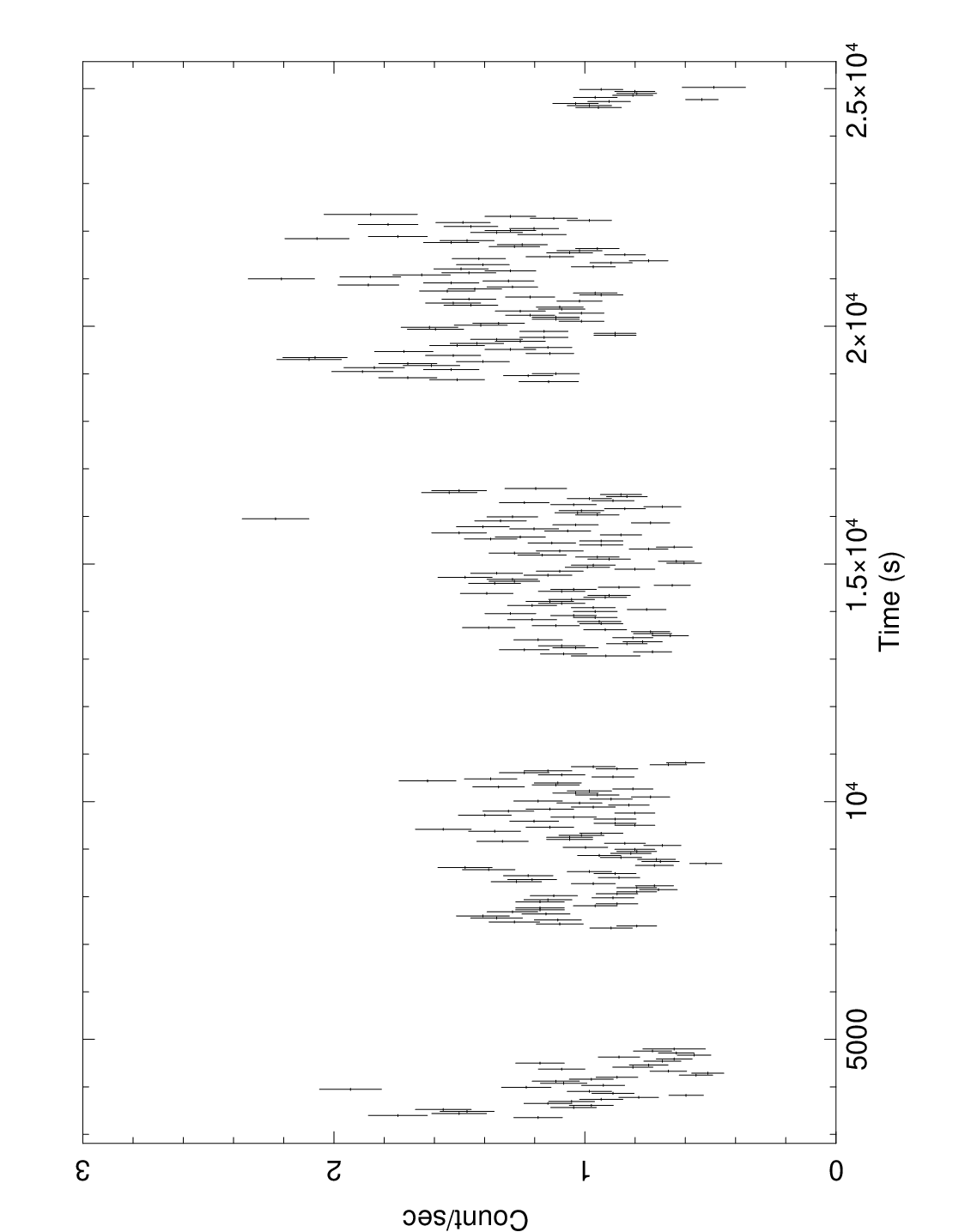}}
\epsscale{0.6} \rotatebox{270}{\plotone{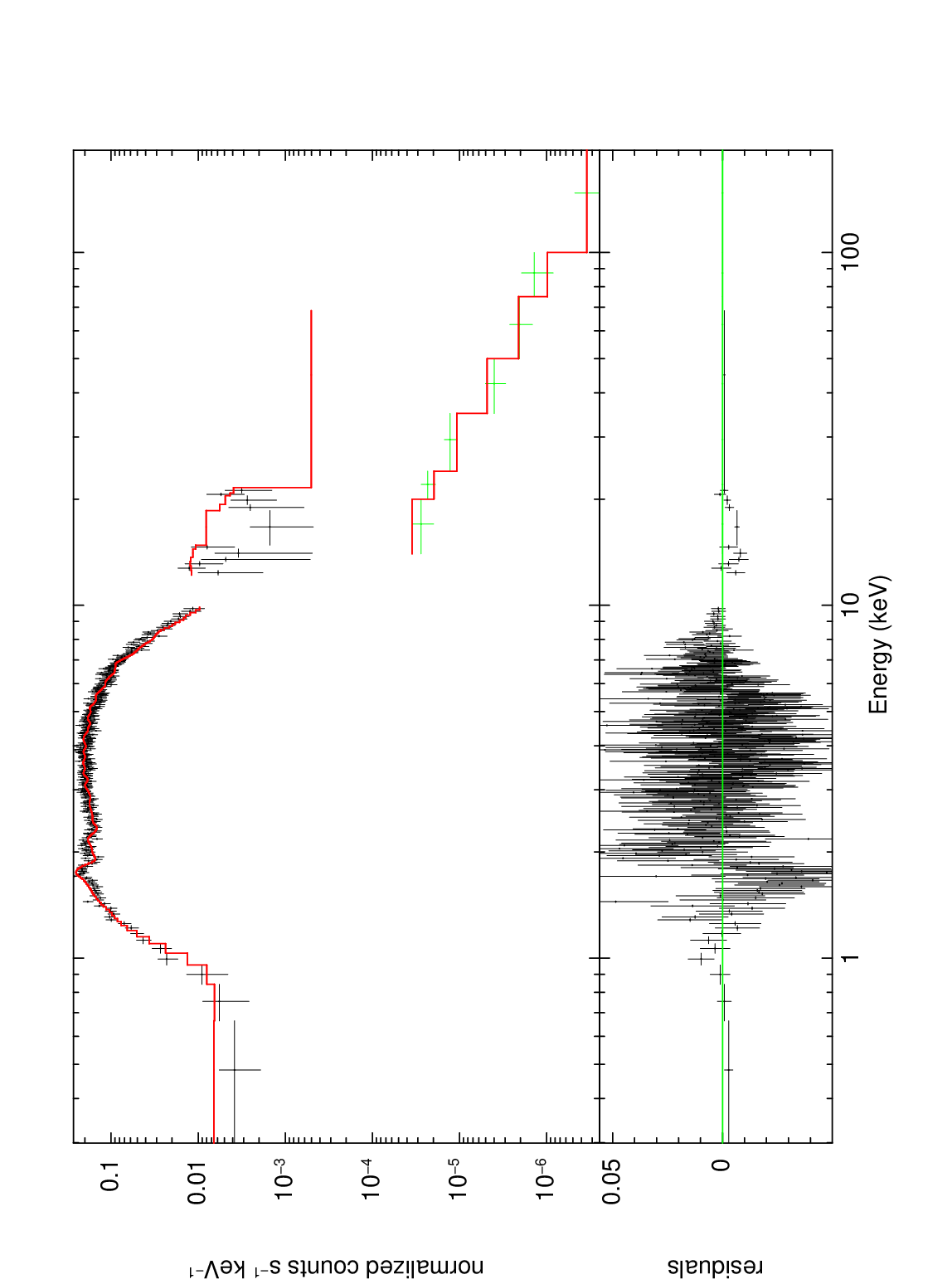}}
\caption[SWJ2000.6+3210 Lightcurve (Top) and spectrum in the 0.2-150.0 keV energy range (Bottom) from Observation 1. Suzaku data are shown in black while BAT data are shown in green. The lightcurve is shown here, where no periodic variability is seen, for comparison to the second observation, where periodicity is apparent.]{\label{figure1}SWJ2000.6+3210 Lightcurve (Top) and spectrum in the 0.2-150.0 keV energy range (Bottom) from Observation 1. Suzaku data are shown in black while BAT data are shown in green. The lightcurve is shown here, where no periodic variability is seen, for comparison to the second observation, where periodicity is apparent.}
\label{fig4}
\end{figure}

\begin{figure}
\figurenum{5} \epsscale{0.4} \rotatebox{270}{\plotone{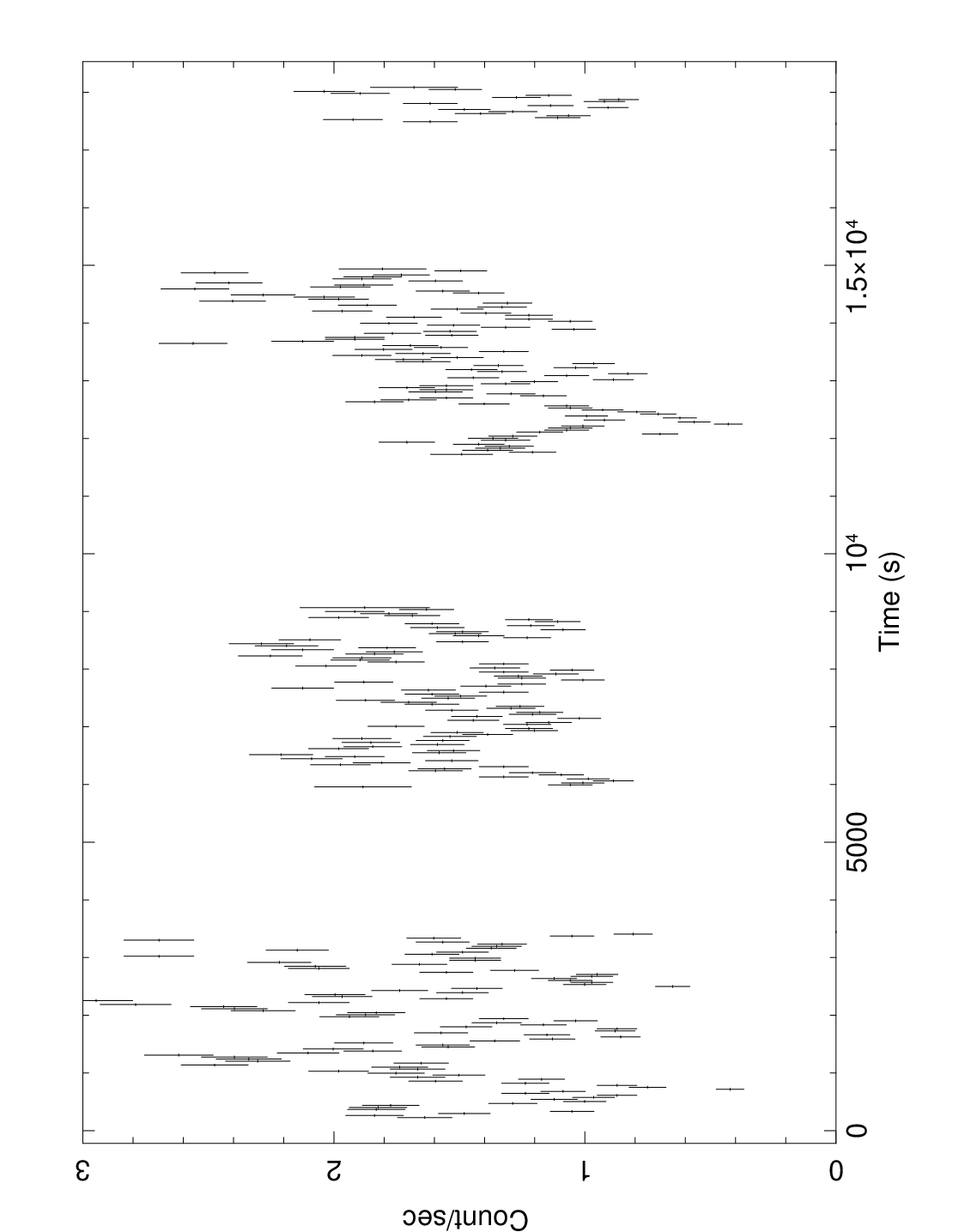}}
\epsscale{0.4} \rotatebox{270}{\plotone{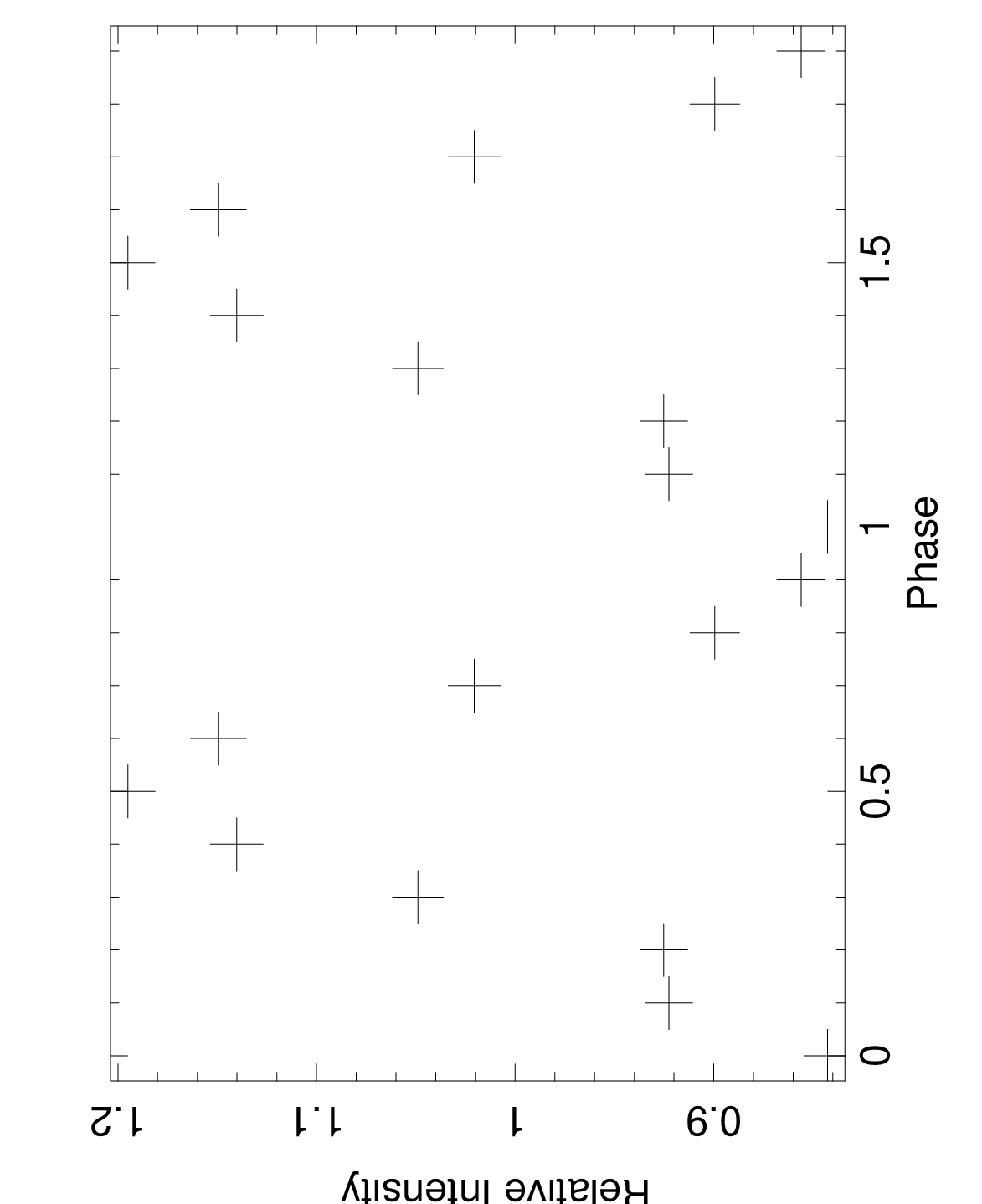}}
\epsscale{0.4} \rotatebox{270}{\plotone{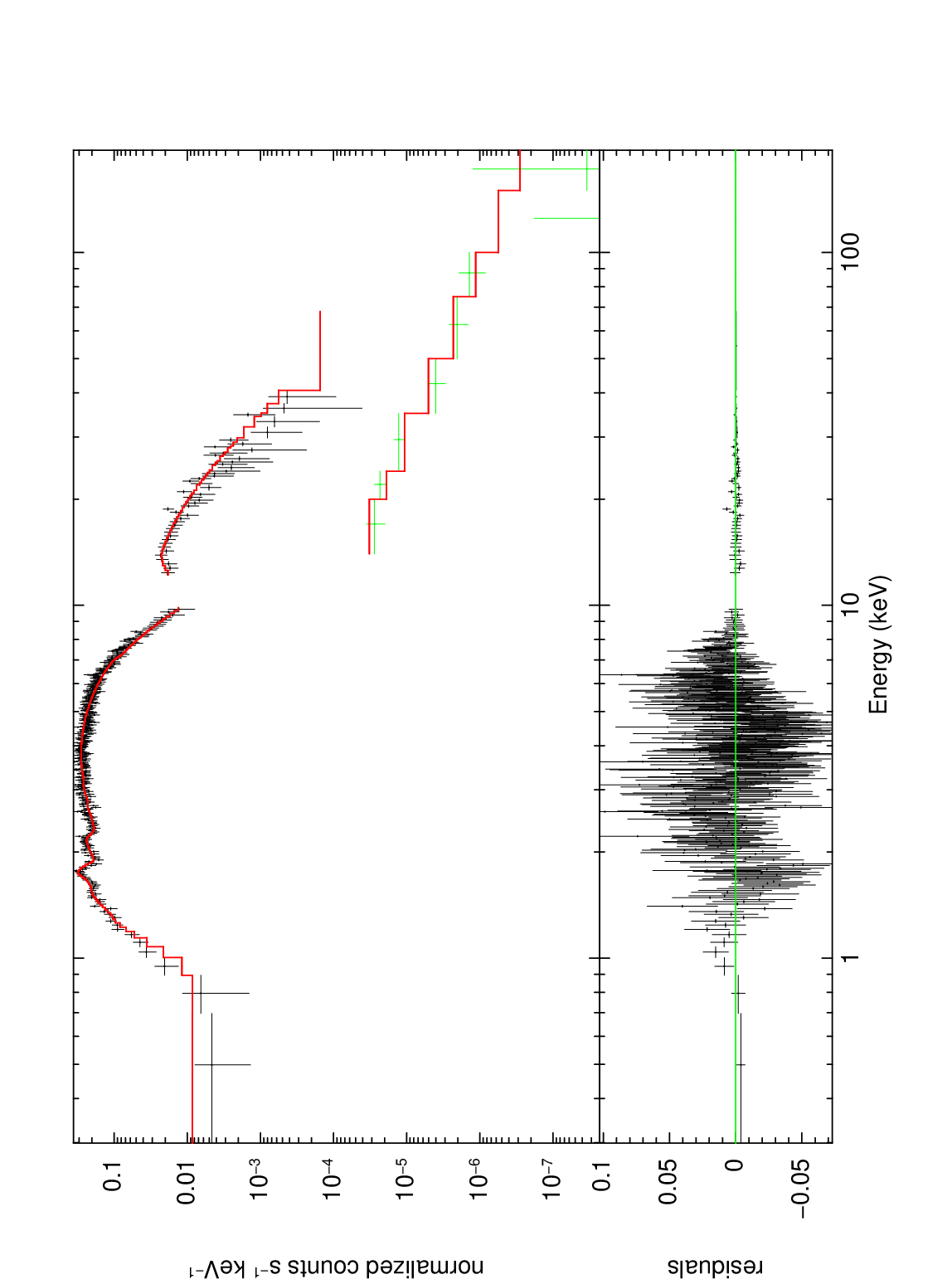}}

\caption[SWJ2000.6+3210 Lightcurve (Top-Left), folded lightcurve (Top-Right) and spectrum in the 0.2-150.0 keV energy range (Bottom) from Observation 2. Suzaku data are shown in black while BAT data are shown in green. Periodicity that may be seen in the raw lightcurve and is made apparent in the folded lightcurve.]{\label{figure1}SWJ2000.6+3210 Lightcurve (Top), folded lightcurve (Middle) and spectrum in the 0.2-150.0 keV energy range (Bottom) from Observation 2. Suzaku data are shown in black while BAT data are shown in green. Periodicity that may be seen in the raw lightcurve is made apparent in the folded lightcurve.}
\label{fig5}
\end{figure}

\end{document}